# A measurement method of transverse light-shift in atomic spin co-magnetometer

Li Xing[1,2,*], Wei Quan[2,3], Tianxiao Song[4], Qingzhong Cai[2] and Wen Ye[5]

[1] *Division of Thermophysics and Process Measurements, National Institute of Metrology, NIM, Beijing 100029, China*
[2] *School of Instrumentation Science and Opto-electronics Engineering, Beihang University, Beijing 100191, China*
[3] *Beijing Academy of Quantum Information Sciences, Beijing 100191, China*
[4] *Beijing Institute of Space Launch Technology, China Academy of Launch Vehicle Technology, Beijing 100076, China*
[5] *Division of Mechanics and Acoustic Metrology, National Institute of Metrology, NIM, Beijing 100029, People's Republic of China*
*Corresponding author: xingli@nim.ac.cn

**Abstract**

We disclose a method to obtain the transverse light-shift along the probe light of a single-axis alkali metal-noble gas co-magnetometer. The relationship between transverse compensating field and light-shift is deduced through the steady-state solution of Bloch equations. The variety of probe light intensity is used to obtain the residual magnetic field, and step modulation tests are applied to acquire the total spin-relaxation rate of electron spins and self-compensation point. Finally, the transverse light-shift is reduced from -0.115 nT to -0.039 nT by optimizing the probe light wavelength, and the value of the calibration coefficient can be increased simultaneously.

## 1. Introduction

An alkali metal-noble gas co-magnetometer has found a wide range of applications in sensing rotation rate [1-3], testing Lorentz and CPT violation [4,5], and searching for anomalous spin-dependent forces [6,7]. In atomic spin co-magnetometers, the electron spins of alkali-metal are in SERF regime [8], so that the electron spins are ultra-high sensitive to magnetic field [9]. The nuclear spins of noble gas can be hyperpolarized by spin-exchange collisions with the polarized electron spins [10]. The polarized nuclear spins have ability of self-compensation to the external magnetic fields, and meanwhile the ultra-high sensitivity of rotation rate can be maintained [2]. However, the light-shift [11,12] arising from circularly polarized component of non-ideal linearly polarized probe beam can affect the direction and stability of electron spins [13,14]. Thus, the research on measurement and optimization of the transverse light-shift is essential for atomic spin co-magnetometers.

Recently, the influence of light-shift caused by the circularly polarized pump light of atomic spin co-magnetometer has been widely studied [15-17]. The light-shift interaction appears as a fictitious magnetic field, so that the atomic spins process around the total effective magnetic field [18]. Therefore, the light-shift should be minimized as far as possible. Moreover, the principle of light-shift caused by the pump light of a K-Rb-21Ne co-magnetometer has been analyzed [19], and the light-shift of Rb atoms can be reduced by collision mixing [11]. Furthermore, the light-shift would lead to cross-talk effect for a dual-axis co-magnetometer, which limits the measurement accuracy of the two sensitive axes [2,20]. However, there are few studies on transverse light-shift caused by the probe light of co-magnetometer, which has usually been ignored in the steady-state solution of Bloch equations [21,22]. With the improvement of the accuracy, the existence of the transverse light-shift cannot be neglected.

In this letter, we propose a method to measure the transverse light-shift based on the steady-state solution of Bloch equations through the variety of probe light intensity and transverse compensation magnetic field. The relationship between transverse light-shift and wavelength of the probe light has been studied, and we propose an optimization method for reducing the light-shift. Finally, the transverse light-shift has been reduced from -0.115 nT to -0.039 nT and the light-shift related term has been effectively suppressed. Meanwhile, the calibration coefficient of co-magnetometer has been concurrently increased.

## 2. Theory of the measurement method

The behavior of the K-Rb-$^{21}$Ne co-magnetometer can be represented by a set of Bloch equations. The evolutions of electron spin polarization $\mathbf{P}^e$ and nuclear spin polarization $\mathbf{P}^n$ can be described as below [1,2],

$$\frac{\partial \mathbf{P}^e}{\partial t} = \frac{\gamma_e}{Q(P^e)}(\mathbf{B}+\mathbf{B}_n+\mathbf{L})\times\mathbf{P}^e - \mathbf{\Omega}\times\mathbf{P}^e + \frac{(R_p\mathbf{s}_p + R_{se}^{en}\mathbf{P}^n + R_m\mathbf{s}_m - R_{tot}^e\mathbf{P}^e)}{Q(P^e)}$$

$$\frac{\partial \mathbf{P}^n}{\partial t} = \gamma_n \left( \mathbf{B} + \mathbf{B}_e \right) \times \mathbf{P}^n - \mathbf{\Omega} \times \mathbf{P}^n + R_{se}^{ne} \left( \mathbf{P}^e - \mathbf{P}^n \right) - R_{sd}^n \mathbf{P}^n \qquad (1)$$

where $\mathbf{\Omega}$ is the rotation velocity. $Q(P^e)$ is the slowing-down factor of electron. $\gamma_e$ and $\gamma_n$ are the gyromagnetic ratios of electron and nuclear spins, respectively. $R_p$ and $R_m$ are the pumping rates produced by the pump and probe lights. $R_{se}^{ne}$ and $R_{se}^{en}$ are the spin-exchange rates from electron to nucleus and from nucleus to electron, respectively. $R_{tot}^e$ is the total spin-relaxation rate of electron spins, which is equal to $R_p + R_m + R_{se}^{en} + R_{sd}^e$. $R_{sd}^e$ and $R_{sd}^n$ are the electron and nuclear spin-destruction rate, respectively. $\mathbf{B}_e$ and $\mathbf{B}_n$ are the magnetic fields generated by the magnetizations of electron and nuclear spins. The electron spins process in the residual magnetic field $\mathbf{B}$. $\mathbf{L}$ is the light shift (AC-Stark shift) field, which is a magnetic-like field coupling to electrons generated by the detuned pump and probe lights. When the probe beam is along x-direction, the steady-state solution of Bloch equations about $L_x$ can be simplified as follows,

$$P_x^e(L_x) \approx \frac{\gamma_e P_z^e R_{tot}^e}{R_{tot}^{e\,2} + \gamma_e^2 \left( L_z + \delta B_z \right)^2} \left[ -\frac{\delta B_z}{B_n} \frac{L_x \gamma_e}{R_{tot}^e} B_c + \frac{\gamma_e L_x L_z}{R_{tot}^e} + o(10^{-6}) \right] \qquad (2)$$

where $\delta B_z = B_z - B_c$ is residual magnetic field along z-axis after the coil compensation. $B_c = -B_n - B_e$ is self-compensation point of the co-magnetometer to cancel the fields from nuclear and electron magnetization. The transverse light-shift $L_x$ are mainly related to $\delta B_z$, and the longitudinal light shift $L_z$. It is approximated to a constant with the parameters of temperature and pump light invariant. Thus, the $\delta B_z$ related terms of transverse electron spin polarization $P_x^e$ in a K-Rb-$^{21}$Ne co-magnetometer can be deduced as,

$$P_x^e(\delta B_z) \approx \frac{\gamma_e P_z^e R_{tot}^e}{R_{tote}^2 + \gamma_e^2 \left( L_z + \delta B_z \right)^2} \left[ -\frac{\delta B_z}{B_n} (B_y + \frac{L_x \gamma_e}{R_{tot}^e} B_c) + o(10^{-6}) \right] \qquad (3)$$

Once the compensation field of $B_z$ has been found, the residual magnetic field $B_y$ can be compensated by the coils. We note that $B_y$ and $L_x$ jointly determine whether the magnetic field along y-axis is zero. The value of transverse compensating magnetic field $B_{yc}$ is equal to $-B_y - \frac{L_x \gamma_e}{R_{tot}^e} B_c$. Thus, when the residual magnetic field $B_y$ is determined, the light shift $L_x$ can be obtained. Considering total energy level splitting effect of the probe light on D1 and D2 lines of Rb atoms, the light shift $L_x$ can be expressed by [11],

$$L_x = \frac{\Phi(v) r_e c}{A \gamma_e} \left( -f_{D1} \frac{(v_{probe} - v_{D1})}{(v_{probe} - v_{D1})^2 + (\Gamma_{D1}/2)^2} + f_{D2} \frac{(v_{probe} - v_{D2})}{(v_{probe} - v_{D2})^2 + (\Gamma_{D2}/2)^2} \right) s_x \qquad (4)$$

where $r_e$ is classical electron radius and $c$ is light velocity. $f_{D1}$ and $f_{D2}$ are oscillator strength of D1 and D2 lines, respectively. $s_x$ is ellipticity of the non-ideal probe light. $v_{D1,D2}$ are the D1 and D2 resonance lines, respectively. $\nu_{probe}$ is the frequency of probe beam. $\Gamma_{D1,D2}$ are broadening width under D1 and D2 resonance lines, respectively. $\Phi(\nu)/A$ is the luminous flux per unit area and per unit time, which is proportional to the light intensity. The $|L_x|$ increases linearly with the increasing of probe intensity, so that the relationship between transverse compensating field $B_{yc}$ and probe intensity are linear. When the light intensity is zero, the light shift $L_x$ can be zeroed at the same time and $B_{yc} = -B_y$. Therefore, the residual magnetic field $B_y$ can be obtained by varying the intensity of probe light and deducing $B_{yc}$ at the point of zero intensity light. Finally, the value of $L_x \gamma_e B_c / R_{tot}^e$ can be acquired by adding $-B_{yc}$ and $-B_y$. Moreover, the total relaxation rate $R_{tot}^e$ and the compensation point $B_c$ can be calculated by fitting the dispersion curves generated by $B_y$ modulation as a function of $\delta B_z$. The step modulation output signals $\Delta S$ can be expressed as follows,

$$\frac{\Delta S}{\Delta B_y} = k \frac{P_e^z R_{tot}^e}{\gamma_e B_c} \cdot \frac{\delta B_z}{((\delta B_z + L_z)^2 + (R_{tot}^e / \gamma_e)^2)} \tag{5}$$

where $k$ is a constant relating to the . Therefore, the transverse light-shift $L_x$ can be calculated by,

$$L_x = -\frac{R_{tot}^e (B_{yc} + B_y)}{\gamma_e B_c} \tag{6}$$

According to Eq. (4), the relationship between probe light wavelength and transverse light shift $L_x$ under different $^{21}$Ne pressure can be simulated in Fig.1. Thus, the $L_x$ can be reduced by optimizing the detuning of probe light wavelength.

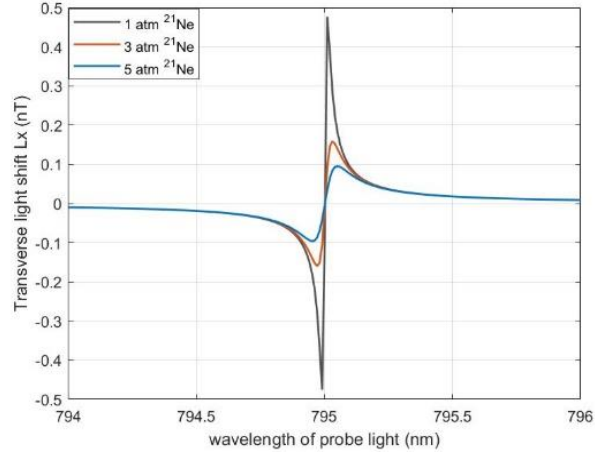

**Fig.1.** The simulation of the relationship between probe light wavelength and $L_x$ under different pressure.

## 3. Experimental setup and results

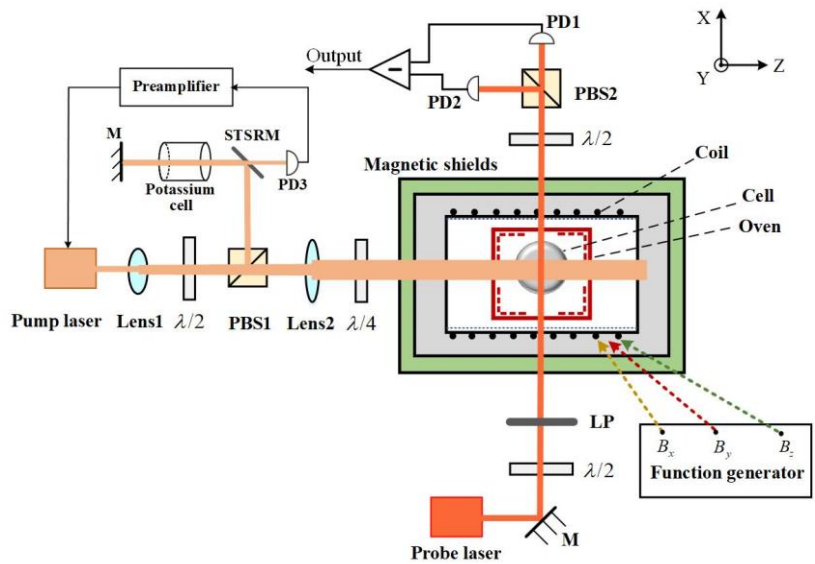

**Fig.2.** Schematic diagram of the K-Rb-$^{21}$Ne co-magnetometer. Lens: plano-convex lens, LP: linear polarizer, M: reflection mirror, PBS: polarization beam splitter, STSRM: semi-transparent and semi-reflective mirror, PD: photodiode.

The schematic diagram of the K-Rb-$^{21}$Ne co-magnetometer is shown in Fig.2. The sensitive core is a 10-mm-diameter spherical cell, which contains a drop of K and Rb alkali metals, 3 atm of $^{21}$Ne and 40 Torr of $N_2$ for quenching. The vapor cell is placed in an oven, which can heat the temperature of the cell to 180 °C. Three layers of $\mu$- metal magnetic shields surround the oven to provide a weak magnetic environment for atoms. A set of three-axis magnetic field coils is used to compensate residual fields in the shields.

A circularly polarized pump light along z-axis tuned on the K D1 resonance line is used to polarize the K atoms, and the Rb atoms are polarized through the spin-exchange collision among K atoms. The polarized alkali-metal atoms ultimately hyperpolarize the $^{21}$Ne atoms [23]. A potassium vacuum cell is used to present the absorption spectrum, and the wavelength of pump light can be stabilized to the order of megahertz by saturated absorption. Thus, the influence of pump light wavelength on light-shift can be ignored in our setup. A non-ideal linearly polarized probe beam detuned to the red side of Rb D2 line, which is orthogonal to the pump beam, is utilized to detect the variety of the transverse polarization of Rb atoms along x-axis. The probe beam contains elliptical polarization component due to the limitation of polarization performance of the polarizer, which can cause transverse light-shift along the probe beam. The relationship between probe light intensity and compensating field $B_{yc}$ is shown in Fig.3. According to the linear fitting curves, the residual magnetic field $B_y$ can be obtained by the intercept. The insert map shows measurement results of $B_y$ under different light wavelength and the average value is 3.379 nT in this setup.

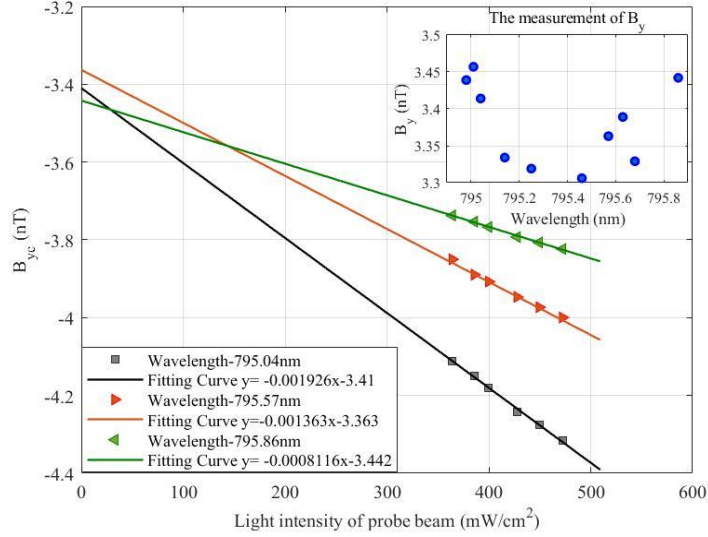

**Fig.3.** Relationship between the compensating field $B_{yc}$ and the probe light intensity under different light wavelength. The residual magnetic field $B_y$ can be obtained by the intercept.

The $B_y$ step modulation output signals under different wavelength of the probe light have been tested in Fig.4. The $R_{tot}^e/\gamma_e$ and $B_c$ can be obtained by fitting the dispersion curves with Eq.(5), and the fitting results under different wavelength are shown in the insert map of Fig.4 and Fig.5, respectively. It is obvious that the values of $R_{tot}^e/\gamma_e$ and $B_c$ are influenced by the wavelength of probe light due to transverse pumping effect on the electron spins. According to the values of $R_{tot}^e/\gamma_e$ and $B_c$, the light-shift $L_x$ can be calculated by Eq.(6). The values of $L_x$ and the related term $\frac{L_x \gamma_e}{R_{tot}^e} B_c$ under different wavelength of probe light are shown in Fig.5. The relationship between $L_x$ and the wavelength of probe light is a Lorentz linear, and the effect of $L_x$ can be magnified by the terms of $R_{tot}^e/\gamma_e$ and $B_c$ for about one order of magnitude. $L_x$ has the most serious effect near the wavelength of 795 nm, which is at the transition frequency of Rb D2 line. When the wavelength is detuned to the red side of Rb D2 line, $L_x$ decays exponentially and it is basically unchanged with the light wavelength larger than 795.6 nm. At 3 atm of $^{21}$Ne, the calibration coefficient of co-magnetometer firstly increases and then decreases with the increase of the light wavelength. The performance of the co-magnetometer can be optimized by reducing the influence of transverse light-shift $L_x$ and meanwhile increasing the calibration coefficient. In our experiments, the $L_x$ can be reduced from -0.115 nT to -0.039 nT by optimizing the wavelength of probe light to 795.68 nm. Meanwhile, the calibration coefficient is 26.19 V/°/s. When the wavelength is greater than 795.68 nm, the value of calibration coefficient gradually decreases and the value of $L_x$ is almost invariant. Thus, the optimal wavelength of the probe light is 795.68 nm for our setup, and the related term $\frac{L_x \gamma_e}{R_{tot}^e} B_c$ can be reduced from 1.113 nT to 0.431 nT during the optimization routine.

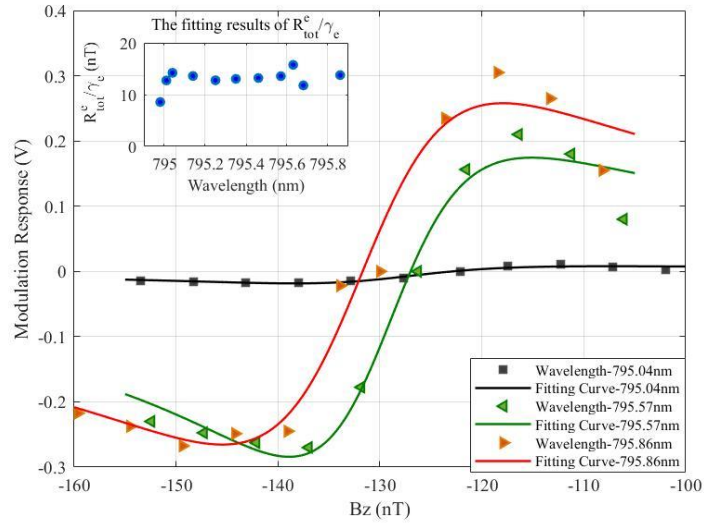

**Fig.4.** $B_y$ step modulation output signals with different wavelength of the probe light. The fitting curves are derived from Eq.(5).

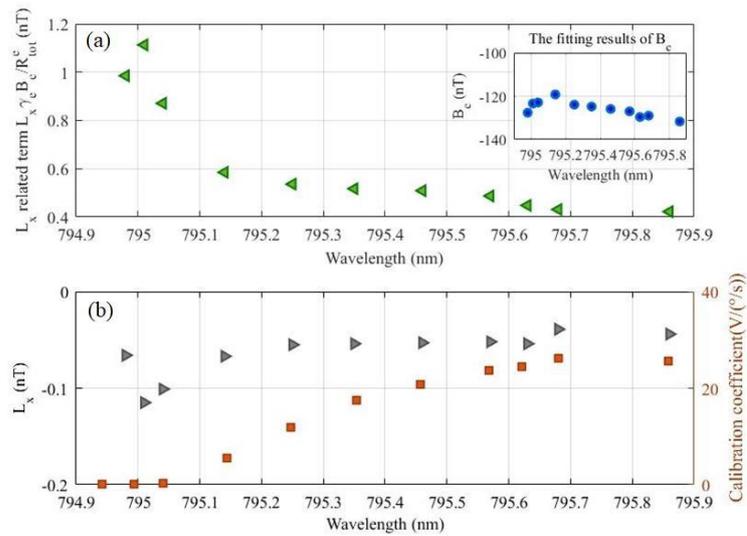

**Fig.5.** Optimization results of $L_x$, the related term $L_x \gamma_e B_c / R_{tot}^e$ and calibration coefficient. (a) The tests of related term $L_x \gamma_e B_c / R_{tot}^e$. The fitting results of $B_c$ is derived from the data of Fig.5. (b) The test results of

$L_x$ and calibration coefficient of the co-magnetometer.

## 4. Conclusion

In conclusion, a method for measuring the transverse light-shift has been examined in a compact K-Rb-$^{21}$Ne co-magnetometer. The value of the transverse light-shift $L_x$ could be calculated by Eq.(6). The residual magnetic field $B_y$ could be obtained by measuring the relationship between transverse compensating field $B_{yc}$ and the intensity of probe light. In addition, other related parameters $R_{tot}^e/\gamma_e$ and $B_c$ could be obtained by fitting the $B_y$ step modulation output signal $\Delta S$. Finally, the transverse light-shift $L_x$ has been reduced from -0.115 nT to -0.039 nT by optimizing wavelength of the probe light to 795.68 nm. Meanwhile, the related term $L_x\gamma_e B_c/R_{tot}^e$ has been reduced from 1.113 nT to 0.431 nT. Therefore, the influence of the transverse light-shift could be effectively restrained, which is beneficial to improve the accuracy of rotation rate measurement. Further improvement may be realized by properly increasing the pressure of $^{21}$Ne in the vapor cell, which can directly inhibit the generation of transverse light-shift.

## 5. Acknowledgement

This work was supported in part by the National Natural Science Foundation of China under Grant 61803015, Grant 61901431. Basic Scientific Research Fund of NIM under AKYJJ1906.

## References

1. T. W. Kornack, R. K. Ghosh, and M. V. Romalis, Phys. Rev. Lett. 95, 230801 (2005).
2. R. Li, W. Fan, L. Jiang, L. Duan, W. Quan, and J. Fang, Phys. Rev. A 94, 032109 (2016).
3. M. Romalis and T. Kornack, (Princeton University Princeton United States, 2018).
4. F. Allmendinger, W. Heil, S. Karpuk, W. Kilian, A. Scharth, U. Schmidt, A. Schnabel, Y. Sobolev, and K. Tullney, Phys. Rev. Lett. 112, 110801 (2014).
5. M. Smiciklas and M. Romalis, (Meeting of the APS Division of Atomic American Physical Society, 2012).
6. L. Hunter, J. Gordon, S. Peck, D. Ang, and J.-F. Lin, Science 339, 928–932 (2013).
7. D. F. J. Kimball, J. Dudley, Y. Li, D. Patel, and J. Valdez, Phys. Rev. D 96, 075004 (2017).
8. J. C. Allred, R. N. Lyman, T. W. Kornack, and M. V. Romalis, Phys. Rev. Lett. 89, 130801 (2002).
9. W. Quan, W. Kai, and R. Li, Opt. Express 25, 130801 (2017).
10. R. K. Ghosh and M. V. Romalis, Phys. Rev. A 81, 043415 (2010).
11. Y. Chen, W. Quan, L. Duan, Y. Lu, L. Jiang, and J. Fang, Phys. Rev. A 94, 052705 (2016).
12. D. Miletic, C. Affolderbach, M. Hasegawa, R. Boudot, C. Gorecki, and G. Mileti, Appl. Phys. B 109, 89–97 (2012).
13. G. Vasilakis, Ph.D. thesis, Princeton University (2011).

14. I. A. Sulai, R. Wyllie, M. Kauer, G. S. Smetana, R. T. Wakai, and T. G. Walker, Opt. letters 38, 974–976 (2013).
15. K. Wei, T. Zhao, J. Fang, Y. Zhai, R. Li, and W. Quan, Opt. Express 27, 974–976 (2019).
16. C. Jia, C. Liu, D. Ming, C. Zhen, Y. Liang, and F. Wu, Appl. Phys. Lett. 116, 142405 (2020).
17. E. Zhivun, A. Wickenbrock, J. Sudyka, B. Patton, S. Pustelny, and D. Budker, Opt. Express 24, 15383–15390 (2016).
18. I. Novikova, A. B. Matsko, V. L. Velichansky, M. O. Scully, and G. R. Welch, Phys. Rev. A 63, 63802–63802 (2001).
19. J. Fang, S. Wan, Y. Chen, and R. Li, Appl. Opt. 51, 7714 (2012).
20. L. Jiang, W. Quan, R. Li, L. Duan, W. Fan, Z. Zhuo, F. Liu, L. Xing, and J. Fang, Phys. Rev. A 95, 062103 (2017).
21. R. Li, W. Quan, W. Fan, L. Xing, and J. Fang, Sensors Actuators A: Phys. 266, 130–134 (2017).
22. K. Wei, T. Zhao, J. Fang, R. Li, Y. Zhai, C. Han, and W. Quan, Phys. Rev. A 13, 044027 (2020).
23. R. K. Ghosh and M. V. Romalis, Phys. Rev. A 81, 043415–043415 (2010).